\documentclass[runningheads]{llncs}
\usepackage{graphicx}
\usepackage{hyperref}
\usepackage{color}

\usepackage{caption}
\usepackage{subcaption}
\usepackage{listings}

\begin{document}
\title{Agent-Oriented Visual Programming for the Web of Things}
%
%
\author{Samuele Burattini \inst{1} \and
Alessandro Ricci \inst{1} \and
Simon Mayer \inst{2} \and
\\
Danai Vachtsevanou \inst{2} \and
Jeremy Lemee \inst{2} \and
Andrei Ciortea \inst{2} \and
Angelo Croatti \inst{1}
}
\authorrunning{S. Burattini et al.}
%
\institute{Dipartimento di Informatica - Scienza e Ingegneria,  \\ Alma Mater Studiorum,  Università di Bologna, Cesena Campus, Italy\\
\email{samuele.burattini@studio.unibo.it, \{a.croatti|a.ricci\}@unibo.it}
\and Interaction- and Communication-based Systems, Institute of Computer Science, \\ University of St. Gallen, Switzerland
\email{\{andrei.ciotea|danai.vachtsevanou|jeremy.lemee|simon.mayer\}@unisg.ch}
}
\maketitle              
\begin{abstract}
In this paper we introduce and discuss an approach for multi-agent-oriented visual programming.
This aims at enabling individuals without programming experience but with knowledge in specific target domains to design and (re)configure autonomous software.
We argue that, compared to procedural programming, it should be simpler for users to create programs when agent abstractions are employed.
The underlying rationale is that these abstractions, and specifically the belief-desire-intention architecture that is aligned with human practical reasoning, match more closely with people's everyday experience in interacting with other agents and artifacts in the real world.
On top of this, we designed and implemented a visual programming system for agents that hides the technicalities of agent-oriented programming using a blocks-based visual development environment that is built on the JaCaMo platform.
To further validate the proposed solution, we integrate the Web of Things (WoT) to let users create autonomous behaviour on top of physical mashups of devices, following the trends in industrial end-user programming. 
Finally, we report on a pilot user study where we verified that novice users are indeed able to make use of this development environment to create multi-agent systems to solve simple automation tasks.

\end{abstract}

\keywords{Agent-Oriented Programming  \and Visual Programming \\
\and Hypermedia-MAS \and Web of Things}
\section{Introduction}

In this paper we introduce and discuss an approach for {\it agent-oriented visual programming}. 
We argue that this is a first and very important step to investigate whether Multi-Agent Oriented Programming (MAOP) actually may enable individuals without experience in programming to create or modify ever more pervasive and ever more autonomous systems in their surroundings, by raising the level of abstraction from procedures or objects to agents, which people are more familiar with from their everyday experience. 
In doing so, we also revisit one of the core tenets of agent-oriented programming: the intentional stance and ascribing mental qualities to systems. Having a visual agent programming language geared towards non-technical users would allow us to validate this assumption in the context of systems engineering.

Our endeavor is furthermore motivated by two concrete issues experienced in an industrial scenario based on the Web of Things (WoT).
The first one is the ever-increasing interest in forms of end-user programming that shall enable not only experienced programmers but ideally domain experts without programming experience to create or modify software systems of different complexity.
The second one is the need to create or modify solutions featuring different degrees of autonomy of software components in performing tasks in a flexible way, dealing with open, dynamic, distributed WoT environments. From this angle, at the same time, the use of semantic-web technologies allows to discover high-level actions at run-time, which promotes the serendipitous creation of applications in such environments -- given a proper level of abstraction for exploiting them. 

Our approach is built on a blocks-based visual programming environment to create Multi-Agent Systems (MAS) that are then executed on top of existing platforms -- namely the JaCaMo\cite{boissier2013jacamo} platform -- and on Yggdrasil, a framework for Hypermedia MAS\cite{ciortea2018yggdrasil} with native support for hypermedia environments.

\section{Background and Related Work}

\subsection{Visual Programming for End-Users}
Visual Programming (VP) is defined as the action of programming using more than one dimension to convey semantics~\cite{burnett1995visual}. Traditional text-based programming is considered mono-dimensional since, although visually organized on a 2D screen, code can be seen as a single string of characters. 
In general, the main goals of VP are to make programming more accessible to some particular audience and to improve the correctness and speed with which people perform programming tasks.

VP has been often paired with the idea of End-User Programming (EUP) and End-User Development (EUD), which can be seen as programming done by someone who is not a programmer in their regular work life~\cite{nardi1993small}. 
This field is highly relevant if we think about the fact that most software (in terms of quantity) is written by these people.
EUP is sometimes also defined depending on the \textit{goal} that the programmer has when writing software rather than his skill level. This is usually to automate a personal task, thus not necessarily taking into consideration all the features that are typical in software engineering such as maintenance or testing~\cite{ko2011state-eup}.

The two most popular examples of Visual Programming Environments applied to end-user (and novice users) are block-based visual programming and flow-based visual programming\cite{mason2017block-vs-flow}.
In block-based programming the core idea is to present the user with a primitives-as-puzzle-pieces metaphor to give users visual cues indicating where and how instructions may be used by dragging-and-dropping them together. Syntax errors are prevented since the programming environments forbids to snap together blocks that shouldn't be connected\cite{weintrop2019block}.
Flow-based abstractions instead see programming as the coordination of parallel flows to transform data. This is achieved through the use of components acting as ``black boxes" that can be joined together to create streams of computation. The focus is on reusing predefined functions and combining them together to achieve the final result\cite{morrison1994flow}.

Both approaches have been proven successful in introducing people to programming spanning a range of different domains, including IoT\cite{ray2017survey} which is the one taken in consideration as the use case for our Visual AOP language as well.
Some popular examples of block based tools are MIT AppInventor\cite{pokress2013app-inventor} and Scratch \cite{maloney2010scratch}. Among of the most famous examples of flow-based programming instead are Node-RED\footnote{\url{https://nodered.org/}} and IFTTT\footnote{\url{https://ifttt.com/docs}}.

With the spreading of automation and, hence, the use of computers, throughout all aspects of our lives, there is an ever-growing gap between {\it domain} expertise and {\it implementation} expertise.
The tasks that are delegated to automation systems are becoming increasingly abstract -- to give an example, in today's building automation the user is not anymore expected to manually control window blinds; rather, the goal of keeping the building in a comfortable state with respect to lighting, temperature, and glare is delegated to the system, and users merely set the parameters of this goal.

Against this backdrop, it is important that EUP is extended to scenarios where domain experts (e.g., building automation systems engineers) are enabled to create and configure the behavior of a system at a higher abstraction level as well -- this extension leads to an extension of the EUP concepts to something akin to \textit{domain-expert programming} (DEP), where the programmer is not necessarily just any final customer of a software solution, but a person with deep knowledge of the domain in which the software may be useful. With this, though, we're not meaning to exclude ``regular end-users" since everybody can be considered the domain expert of his own home.

We argue that by empowering such experts with visual tools it would be possible to reduce this gap.  Due to the aforementioned higher abstraction level, it is at the same time highly compelling to build such DEP systems not on the basis of paradigms that traditionally underlie EUP programming (e.g., procedural, flow-based, or object-oriented programming) but to instead use MAOP and its abstractions as the foundation of our system.

\subsection{Related Agent-Oriented approaches}

The idea presented in this paper is strongly related to works in literature that investigate high-level approaches to agent-oriented programming, that have been proposed since the very beginning of agent programming literature.
%



 In~\cite{10.1145/238386.238568}, authors proposed an end-user programming system enabling users to program the behaviour of their personal software agent using rules for their agents to follow. 
 This work can be considered a specific example of approaches that explored  programming by demonstration or programming by example~\cite{10.5555/369505}, investigating a different way to think about programming, towards a perspective in which the programmer is more a teacher instructing an agent what to do.
 A main seminal example in this case is the KidSim environment~\cite{10.1145/223904.223908}, designed for children, to support their creative constructions of simulations to learn.

 In the related context of teaching novice programming~\cite{10.1145/330534.330544}, \cite{min12} proposed a visual programming language/environment (an extension of Scratch called BYOB, today known as Snap!\footnote{\url{https://snap.berkeley.edu}}) makes it possible to specify goals and plans using a block-based visual notation.

\subsection{Hypermedia Environments, the WoT, and Web-based MAS}


The World Wide Web is a distributed hypermedia system. Roy T. Fielding, who led the development of the architectural style of the Web~\cite{rest}, defined hypermedia as ``the simultaneous  presentation of information and controls such that the information becomes the affordance through which the user obtains choices and selects actions''\footnote{Roy T. Fielding, A Little REST and Relaxantion, ApacheCon 2008: \url{https://roy.gbiv.com/talks/200804_REST_ApacheCon.pdf}}.
A hypermedia environment can then be understood as a virtual environment that provides hypermedia affordances to the user---and, through these affordances, allows the user to explore and exploit the environment in order to achieve their objectives. This design rationale is captured by the \textit{Hypermedia As The Engine of Application State (HATEOAS)} principle defined by the REST architectural style~\cite{rest}---and is a core tenet of the architecture of the Web: it helped to reduce coupling between Web components, which in turn allowed the Web to scale up to the size of the Internet. 

The HATEOAS principle and the use of hypermedia affordances was picked up, among other initiatives, by the \textit{World Wide Web Consortium (W3C)} standardization efforts for the Web of Things (WoT) to reduce coupling and to promote interoperability in the Internet of Things (IoT) by hiding the protocols and interfaces used to access IoT devices behind abstract interaction patterns and hypermedia affordances. Central to the approach is the W3C Recommendation for the WoT Thing Description (TD)~\footnote{\url{https://www.w3.org/TR/wot-thing-description/}}, which can be used to create machine-readable descriptions of device interfaces and services.
The WoT TD defines three types of so-called {\it Interaction Affordances} that can be exposed by a \textit{Thing} (or \textit{Web Thing})
\begin{itemize}
    \item \textbf{Property Affordances} expose the state of the Thing (e.g., the on/off state of a lamp); 
    \item \textbf{Action Affordances} allow invoking a function of the Thing, which manipulates state (e.g., toggling a lamp on or off) or triggers a process on the Thing (e.g., dim a lamp over time);
    \item \textbf{Event Affordances} describe an event source that asynchronously pushes event data to consumers (e.g., overheating alerts).
\end{itemize}


The HATEOAS principle and the design rationale behind the Web architecture have also been applied to the engineering of Web-based MAS---to design \textit{Hypermedia MAS} (e.g., see~\cite{ciortea2018yggdrasil,mams}). In~\cite{ciortea2018yggdrasil}, the authors introduce Yggdrasil, a platform for Hypermedia MAS that follows the Agents \& Artifacts (A\&A) meta-model~\cite{agentartifact}. A\&A provides a direct conceptual bridge to the W3C WoT through the environment dimension and, in particular, the artifact model, whose interface is defined in terms of observable properties, observable events, and operations. Yggdrasil can thus be used to bring MAS to hypermedia environments and, in particular, WoT environments.


\section{Agent-Oriented Domain-Expert Programming}

%
%
%
%
%

%
%
With Agent-Oriented DEP we refer to activities and tools that allow domain experts -- individuals with domain expertise but who are not professional software developers -- to program multi-agent systems.
%


One the most popular end-user development tool in the mainstream~\cite{ko2011state-eup}, {\it spreadsheets}, allows relatively un-sophisticated users to write programs that represent complex data models, while shielding them from the need to learn programming languages and reducing the potential for errors by constraining users to a specific environment and the possibilities this affords to them. In a similar way, Agent-Oriented DEP should allow its users to implement MAS, shielding them as much as possible from the need to learn the technical aspects and mechanisms that concern traditional AOP languages.

In this paper we focus in particular BDI-based AOP: The BDI model already provides a level of abstraction which could be considered strongly human-oriented. 
Its definition, motivated by the need of having a resource-bounded intelligent agent capable of both means-end planning and weighting of competing alternatives\cite{bratman1988bdi}, was inspired by \textit{human practical reasoning}. To reduce the time spent on deliberation, the model introduced the idea of \textit{intentions}, alongside beliefs and desires, to indicate that, once chosen, an agent should commit to an intention to some degree instead of continuously reconsidering all the possible routes -- similar to how people behave.

We argue that it is possible to exploit this alignment between human practical reasoning and the BDI architecture to create DEP systems that people can effectively use to program MAS since they more closely match our everyday experience in interacting with artifacts and other agents. Then, the hurdle that remains to be cleared is how to design a DEP interface that could reinforce the alignment of agent and human reasoning, hiding the technicalities instead.

%
%

Based on these hypotheses, we have created a visual tool with a blocks-based abstraction that we believe takes an important first step in this direction to support domain experts when programming MAS.

\subsection{A Case Study: WoT Environments managed by Agents}

This exploration was carried out in the context of the European project IntellIoT\footnote{\url{https://intelliot.eu/}} aimed at shaping the next generation of intelligent IoT systems.

The goals of the project served as an ideal use case to apply our idea of Agent-Oriented DEP since the core objectives are to develop autonomous IoT systems while keeping humans ``in the loop''.
There have been other works already in the direction of providing a simple visual tool for users to configure their smart systems (as in \cite{bak2018smart} and \cite{mayer2014configuration}). Our system, though, directly empowers domain experts to program MAS using agent-oriented abstractions.

WoT technologies are used to connect devices and let them be operated seamlessly from the agent point of view following the principles of Hypermedia MAS.
For this reason, not only the visual language was designed and implemented, but also a Runtime Environment integrating multi-agent systems and the WoT is proposed and used in the evaluation step of this project.


\section{Designing a Visual IDE for Agents and WoT}

\subsection{Choice of Visual Abstraction}

The first step in designing our visual language was choosing the most appropriate abstraction to work with.
Three factors were taken into consideration:
\begin{itemize}
    \item Ease of use for individuals without programming experience;
    \item Coherence with the agent paradigm;
    \item Development support.
\end{itemize}

Blocks-based programming and flow-based programming were taken into consideration when analyzing the possible abstractions of choice.
%
%
%
Being the agent paradigm behaviour oriented, and since BDI agents often implemented with the Procedural Reasoning (PRS) System\cite{georgeff1987prs} which is based on the fact that the agent doesn't need to plan since it's equipped with a \textit{plan library} a blocks-based language was seen as the best option. The idea was to have a separate ``chunk'' of blocks for each plan similarly to what has been done by the MIT App Inventor\cite{pokress2013app-inventor} project to model event handling.

Also, the App Inventor project was originally backed by Google which is still supporting a JavaScript library to create, customize and manage blocks on a canvas named \textit{Blockly}\footnote{\url{https://developers.google.com/blockly}}. This is a strong base model on top of which lots of custom other block languages have been developed, and forms an ideal basis for creating our agent-oriented visual language.

\subsection{Reference Syntax and Constructs}

To begin the design of the visual language, it was necessary to study and understand the syntax and identify the individual ``building blocks'' of the agent-oriented programming language chosen as a reference: \textit{Jason}~\cite{jason}, an extended version and implementation of the conceptual language \textit{AgentSpeak}~\cite{rao1996agentspeak}. This language was selected because it is among the most popular agent-oriented languages also due to the support given by the \textit{JaCaMo} platform.

By looking at the language grammar~\cite{bordini2007jason-paper} some observations were made: An agent is composed from an \textit{initialization} section where the programmer can establish the knowledge that the agent has since the beginning of its execution by defining a set of beliefs, a set of goals to pursue, and a set of deductive rules that can be used to simplify the checking of logic conditions.

After the initialization, the programmer can define a \textit{plan library}, where all the procedural knowledge of the agent is stored. This defines what the agent can do and eventually how to handle failure.

From a syntactical point of view, beliefs and goals are represented as logic predicates that can have zero or more terms. A term can be either:
\begin{itemize}
    \item an \textit{atom} which is any lowercase string with no spaces;
    \item a \textit{string} which is some text surrounded by quotes;
    \item a \textit{number} which is any integer or floating-point number;
    \item a \textit{variable} (only in plans) which is identified by a name starting with an uppercase letter and no spaces;
    \item or lists, arithmetic formulas, or even other predicates.
\end{itemize}  

Plans are defined by their \textit{triggering event} -- agent goal changes or agent belief changes -- an optional \textit{context}, which is a logic expression, and the plan \textit{body} which is a sequence of actions that the agent can perform in its environment.
These considerations about the language structure were used to design the language following the following set of principles:

\begin{itemize}
    \item \textbf{Single Responsibility}: Each block should map only one concept of the language. However, to avoid an explosion in the number of blocks, blocks with similar functionality are grouped together in a single block.
    \item \textbf{Composition}: Individual blocks with single responsibilities should be composable to create constructs that implement more complex functionality.
    \item \textbf{Convey Semantics}: Blocks should be designed to let users understand easily what makes them unique and how they can be combined with other blocks.
\end{itemize}

\subsection{Components and Architecture}

During the design process it was necessary to identify all the software components needed to realize the requirements for the system. This was done trying to understand and separate the responsibilities, to maintain a clean, and expandable, architecture. In total, six modules were identified to compose the system:

\begin{itemize}
    \item \textbf{Smart Environment TD Repository} This module is responsible for storing and serving TDs of all available things in the environment. This is necessary to provide access to the TDs for domain experts to be able to better understand them, try them out manually, and utilize the corresponding blocks in the agent code.
    
    \item \textbf{Thing Explorer} This module is an interface to the TD Repository, it provides an easy way for users to test the affordances masking the details of composing the right request to the Thing itself behind forms and buttons generated from the description. 
    
    \item \textbf{Web IDE} This module allows the user to program agents through a visual language and submit them for execution.
    
    \item \textbf{Storage Manager} This module is responsible for the persistence of the user-created agents' code and of the designed run-time configurations that specify which agents should be run together in a MAS. 
    
    \item \textbf{Runtime Environment} This module executes a MAS given the generated source code and a run-time configuration specifying which and how many agents need to exist in the system. The environment is also extended to natively support agent-to-thing interactions.
    
    \item \textbf{Runtime Orchestrator} This module is an optional module that schedules the execution of run-time configurations within different Runtime Environments. Although this was not in the requirements when it came to planning the support for execution of MAS, the possibility of having multiple separated systems running was considered an interesting feature to add.
\end{itemize}

\begin{figure}[t]
    \centering
    \includegraphics[width=\textwidth]{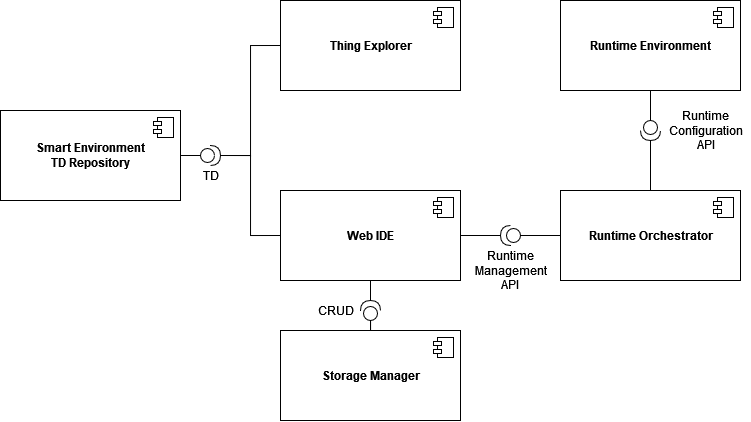}
    \caption{Components and their relationships to implement the systems' requirements}
    \label{fig:component-diagram}
\end{figure}

As shown in Fig.~\ref{fig:component-diagram}, these components collaborate to implement the requirements: The \textbf{Smart Environment TD Repository} exposes an interface to serve TDs to the other components. The \textbf{Thing Explorer} consumes the descriptions to generate a user-friendly interface, whereas the \textbf{Web IDE} does the same to provide programming constructs that can be inserted in the agent visual language. The \textbf{Web IDE} persists the defined agents through CRUD\footnote{Create, Read, Update, Delete} operations provided by the \textbf{Storage Manager} and interacts with the \textbf{Runtime Orchestrator} to schedule the execution of MAS using the \textbf{Runtime Environment} as execution platform.

\section{Prototype Implementation} 

\subsection{Creating the Block Language}

Blocks have been developed based on the \textit{Blockly} framework. Starting from the analysis of the Jason language and following the principles stated above, the process of defining blocks involved specifying the shape and connections that each block should have to allow syntactically correct combinations. 

Blocks programs are transformed to code through the Blockly code generator that allows to implement for each type of block a custom serialization method in a way that resembles classical parsers with the parsing phase already done by the block structure itself.
When defining blocks it is possible to include so-called \textit{mutations}, which are values embedded in the block to keep extra information that is not necessarily visible in the graphic representation. This mechanism was heavily used to generate affordances blocks and store values needed in the code generation phase without showing the users technical details such as affordances URIs.

In total, six categories were defined to group blocks:

\begin{itemize}
    \item The \textbf{Values} category contains blocks for raw values that can be used in combination with other blocks such as atoms, strings, numbers, booleans and variables;
    \item The \textbf{Operations} category comprises operations (e.g., math and logic) with Values blocks as operands;
    \item Blocks from the \textbf{Initialization} category are used to define the initial knowledge of the agent in the form of beliefs, goals, and rules;
    \item \textbf{Plan definition} blocks are used to define plans given their triggering events and context;
    \item \textbf{Agent actions} blocks represent all the actions available to be used in plans and can be extended to support more of the internal actions that Jason supports;
    \item A \textbf{Communication} category is also defined for blocks related to agent-to-agent communication. In Jason communication is expressed by the \texttt{.send} Internal Action that requires to specify alongside the content of the message the intended receiver and the performative as well. At the time being, only the \texttt{tell} (that translate in belief sharing) and \texttt{achieve} (that translate in goal delegation) performatives are actually implemented in block form because they were the easier to convey naturally, but all the others can be easily added in a future expansion of the visual language.
\end{itemize}

Additionally, a category is dynamically created based on the TDs of each smart thing in a workspace to easily group and distinguish blocks related to the use of affordances on different things.

\begin{figure}[h]
    \begin{subfigure}[b]{\textwidth}
        \centering
        \includegraphics[width=0.6\textwidth]{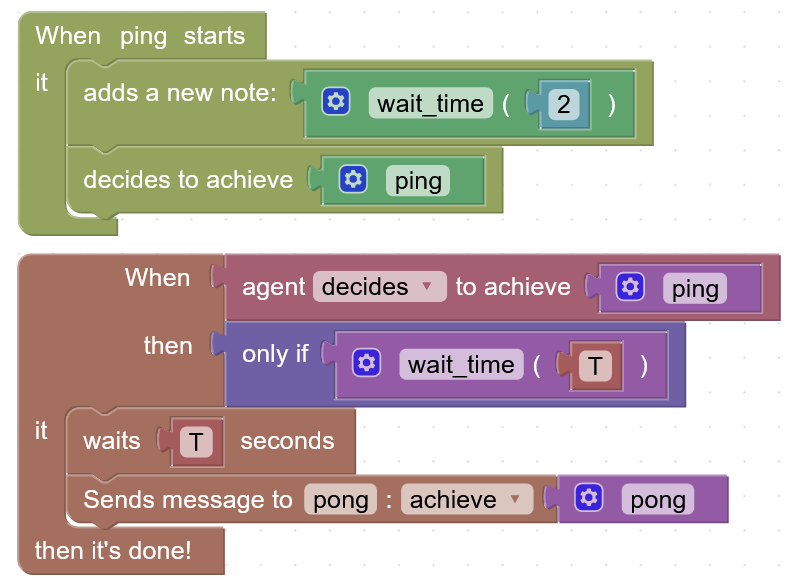}
        \caption{A ``ping'' agent implemented with the block language.}
        \label{fig:ping-agent}
        \vspace{1em}
    \end{subfigure}
    \begin{subfigure}[b]{\textwidth}
\begin{lstlisting}
//This is the initial state of agent ping
wait_time(2).
!ping.

//Plan library:
+!ping : wait_time(T)
  <- .wait(T*1000);
     .send(pong, achieve, pong).
\end{lstlisting}
        \caption{The Jason code generated from the blocks above}
        \label{lst:ping_code}
    \end{subfigure}
    \label{fig:code_and_blocks}
    \caption{An example of the block language with the generated Jason code}
    \vspace{-2em}
\end{figure}

Fig.~\ref{fig:ping-agent} shows several of the blocks available in the block language. This is an implementation of a simple agent with a plan that asks another agent to achieve a ``pong'' goal that is expected to ask the sender to ``ping'' again creating back and forth communication. The initialization of the agent (green blocks) is easily distinguishable from the plan definition (brown). In the initialization, a goal is assigned to the agent and the plan is defined with a matching trigger. The plan is defined with a context that needs to be checked at run time, in this case the plan will be chosen \textit{only if} the agent has a belief (note) telling it how long to wait before sending a new message. 

We have created block labels so as to convey to users that they are instructing agent ``colleagues'' on how they should behave, and have used colors to make it easier to understand which blocks fit together. This is visible in the comparison with the generated Jason code (Fig. \ref{lst:ping_code}) which is definitely less readable for unexperienced users.

\subsection{Smart Environment TD Repository}

We used the Yggdrasil infrastructure to enable our system to discover TDs at run time. However, as this discovery functionality is only one of Yggdrasil's features that is rather generic, our implementation does not depend on this specific infrastructure. In fact, any service that is capable of exposing TDs can be used by our system as long as it organizes TDs according to the Agents \& Artifacts metamodel's environments and workspaces.

Since Yggdrasil can also run digital artifacts remotely, it can integrate both Things and Artifacts seamlessly from the point of view of the agent, using the TD model. This means that agents can also use any other external service or even coordinate through artifacts.

\subsection{Building the Runtime Environment wrapping JaCaMo}

Through the visual language, users can build agents and generate valid Jason source code that could be exported and used in any Jason application. The goal of the project though was to not only support people in the building phase but also in the execution phase like any modern IDE would do. 
The JaCaMo MAS platform~\cite{boissier2013jacamo} has been used to support the execution of the agents.
It is worth remarking that JaCaMo allows for developing and running MAS that can exploit also the environment and organisation as first-class abstractions.
In this work we considered only the agent dimension---nevertheless, the extension of the work to include also the environment and organisation is part of future work.

The visual language is also specific for interaction with the WoT and thus requires a run-time environment properly configured and equipped with the enabling tools.
For these reasons a proper Runtime Environment was required, which is implemented as a Web Server that exposes REST APIs to receive both agent source code and run-time configurations to start the execution and to later stop it at any given moment. The idea of having a Web server comes from the fact that our agent-oriented DEP system is Web-based so it makes sense to have the possibility to directly submit the coded agents to an execution node. 

The integration of JaCaMo with REST has been already studied by the original creators of the platform in a recent paper~\cite{amaral2020jacamo-rest} that led to the creation of JaCaMo-REST: a resource-oriented Web-based abstraction for the multi-agent programming platform implemented in Java.
In its essence, JaCaMo-REST is a JaCaMo application that directly exposes APIs to modify all the relevant entities in a MAS from agents to artifacts and organizations. The idea at the core level is to have a MAS always running and allow an external application to edit which agents are present, the environment and even the behaviour of agents by communicating with them or injecting new plans.

Since the requirements of the Runtime Environment planned for the system were different, JaCaMo-REST was wrapped and slightly modified to work as needed.
Agent source code and configurations are received through the API, converted to the right kind of files to work with the JaCaMo-REST instance. The latter is executed as a subprocess and is started whenever a new configuration is posted.
Communication with the running instance of JaCaMo-REST is achieved using its original APIs over the local network. The only difference with the original implementation is that, to accept the dynamic creation of new agents in a running MAS, the name of the source file for the new agent is passed as a parameter in the API since the original implementation always deployed agents based on an empty template instead.

\subsection{WoT integration in JaCaMo}

Both Jason and JaCaMo were built knowing that, depending on the specific application, developers might have needed additional control over what agents can achieve especially when interacting with external services. This can be done either by creating new internal actions of the Jason language using Java or by defining CArtAgO artifacts~\cite{agentartifact} (once again in Java) that agents can use.

The two methods, although quite similar in practice, have a substantial semantic difference that was considered when choosing how to support proper interaction with Web Things. Actions can be seen as something that an agent can do, they are \textit{internal} by definition thus they are not shared between agents. Artifacts, on the other hand, can be seen as tools providing functionalities to an agent, they have a life cycle, internal state and they can be shared among agents in the same environment.

For these reasons both methods were used to realize the integration with the WoT. In particular, a \texttt{WoTHttpClientArtifact} was defined providing functionalities for agents to invoke affordances on things and retrieve the result as a JSON object. Internal actions were used instead to add JSON parsing and assembling capabilities to agents.

\section{Initial Evaluation}

After developing a working draft of the entire system, in-house testing proved that the solution was robust enough to allow new users to test it out in a controlled environment. A qualitative user study was conducted to understand if the implemented visual language could actually empower people with little or no experience in programming to implement agent-based solutions. 

For the first evaluation, we let 20 users with no common background approach simple and understandable domains and things that are familiar from everyday experience. Participants for the study were recruited by word-of-mouth and thanks to the help of the Behavioural Lab\footnote{\url{https://behaviorallab.unisg.ch/en}} of the University of St. Gallen. In Table \ref{tab:participants} the most relevant demographic features are shown.

Participants were asked to complete several specific tasks with our DEP system and it was measured how long it took each user to complete these tasks individually within a one hour and a half maximum time-slot.
An audio recording was also kept on since it was interesting to apply the \textit{think-aloud protocol} that is commonly used for usability testing in order to follow the user thought process and be able to understand better what they struggled with. To further evaluate more objectively the usability of our system, the System Usability Scale (SUS)~\cite{brooke1996sus} was chosen for an after-study survey.

Since there was interest in understanding how easily individuals with no experience could grasp the concepts of agent-oriented programming, very little training was given to users: A one-page text and a three-minute tutorial video presented agent basics together with the DEP user interface and its features before participants started working on the study tasks.

The results of our initial study are promising: Twelve participants out of twenty managed to complete three out of five tasks. A lot of time was spent on the first task to enter in the correct mindset and understand blocks behaviour.
This is reflected by the difficulty score that participants attributed to task one and its average completion time which was significantly higher than that of the following tasks.
This was underestimated and thus a lot of people ran out of time before attempting the last task. 
Generally, blocks-based interfaces are intuitive but are purposefully built in a way that lets people play and explore the possible combinations which, for people that have never used anything similar, can take some time to get used to. Users also struggled with understanding the idea of interacting through W3C WoT affordances, often confusing properties and actions.

The hardest agent programming concepts to grasp were loops and flow control, especially for people who had some experience with other programming languages and -- when using our DEP system -- did not find what they were used to since the agent paradigm is significantly different from procedural programming. Users also failed in making use of beliefs to simplify solutions which poses the question on how to suggest better how BDI agent minds works and how to exploit these abstractions.

Overall, the system was considered usable receiving an average score on the System Usability Scale of 73.3 out of 100 which is considered above average. An analysis of the participants' reasoning processes showed that they were able to understand the general flow and, with a little more time and guidance, might have been able to solve even more complex problems.

\renewcommand{\arraystretch}{1.1}
\begin{table}
    \centering
    \begin{tabular}{l  l}
    \hline\hline
    \multicolumn{2}{c}{\textbf{Study Participants}}\\
    \hline
    \hline
    \textbf{Total} & 20\\
    \hline
    \textbf{Age} & 11 under 25\\ & 8 between 25 and 50 \\& 1 50 or older\\ 
    \hline
    \textbf{Gender} & 11 female, 9 male\\ 
    \hline
    \textbf{Level of Schooling} & 8 high-school or lower\\ & 4 bachelor's degree \\&  8 master degree or above \\
    \hline
    \textbf{Current Occupation} & 13 students\\ & 5 working students\\& 2 workers \\
    \hline
    \textbf{Programming Experience} & 8 Yes, 12 No \\
    \hline
    \end{tabular}
    \vspace{1em}
    \caption{Demographic data of the study participants}
    \label{tab:participants}
\end{table}

\section{Future Directions}


Our case study has been helpful both to have a first positive evaluation of the value of agent-oriented visual programming and to identify issues and insights useful for further developing the idea and the tools.
%
%
%
%
The development of the visual language is an ongoing process that requires many iterations to be able to pinpoint the basic concepts of agent programming in such a way to make it easily understandable for any user, finding the right metaphors that convey the agent behaviour and empower people to program them efficiently.

%
%
In this paper, BDI has been taken as reference agent programming model/architecture. Nevertheless, the research exploration about agent-oriented visual programming languages is not meant to be be limited to a specific agent architecture or language. For instance, the exploration appears valuable also for recent practical agent programming languages or frameworks that are not BDI-based (main examples include SARL~\cite{6928174}, JADEL~\cite{BERGENTI2017142}, JIAC~\cite{Hirsch2009}, GAML~\cite{10.1007/978-3-642-25920-3_17}).

%
%
%
%
%

The current version of our DEP system is focusing exclusively on the agent dimension of the developed MAS.
Developing an understandable visualization of BDI agents code can help changing not only the interface on which agents are programmed but the whole process as well---introducing either features like programming by example or even ``human-supported automatic planning'' in which the system proposes a plan that can be then edited by the human before approval~\cite{10.5555/3237383.3237504}.
In addition, in previous work, we have also explored how the organization dimension in MAS can provide high-level abstractions that allow, for instance, production engineers to repurpose manufacturing lines on-the-fly through an intuitive Web-based front-end~\cite{10.5555/3237383.3237504}. We expect it would be essential to include organizational specifications and environment configuration as well in our DEP system to fully support users in building complex MAS.
%
%
%

The agent visual programming language proposed in this paper features a level of abstraction which is similar to the corresponding non visual counterpart (i.e., Jason). 
Actually, this could be just the starting point for identifying (visual) languages featuring a higher level of abstraction,  towards models reducing the conceptual gap with respect to e.g. the application domain, like in the case of domain specific languages. 
To this purpose, it will be interesting to explore the relationships with existing research works in Agento Oriented Software Engineering (AOSE) about model-driven approaches and agent-based methodologies~\cite{aose04}, in particular those proposing diagrams for specifying the structure and behaviour of agent-based systems. 

%
%
Finally, the current blocks-based visual language follows quite strictly traditional AOP, in which the agent program is fully defined by the developer and executed by an agent interpreter.  
Actually, the constraint of considering end-users instead of developers triggers the exploration of a different perspective about agent programming itself, turning it more into a form of an interactive communication in which the end-user would \emph{instructs} or \emph{teach} the agent what to do (possibly omitting how to do it), like in the programming-by-example or programming-by-demonstration cases~\cite{10.5555/369505} or in task instructable agents~\cite{10.5555/1622620.1622629}.
Therefore, the investigation of Agent-Oriented Visual Programming in this direction will call for deeply rethinking to cognitive agent programming and cognitive agent architectures, into directions that explicitly support learning as basic core agent capability~\cite{ar2022}.

\section{Concluding Remarks}
The main contribution of this work is the novel application of visual programming techniques to the agent-oriented paradigm in a way that fixes as the ultimate goal the full implementation by novice programmers of multi-agent-based solutions to problems in different WoT powered domains they are experts in. 

This project did not only conceptualize a visual agent language but also the vision for an accessible Integrated Development Environment mixing agent-oriented programming and the Web of Things in a seamless interface for both humans and software agents. It then also developed a prototypal implementation of such a system that was evaluated on a sample of users with promising results to validate the initial assumptions made when defining the system requirements.

Overall this project brought to the realization of a usable tool and more importantly to the exploration of new routes to integrate agent-oriented software engineering with end-user programming in order to enable easy configuration and make so that humans have the appropriate tools to be always in control of ever-increasingly complex systems that exhibits different degrees of autonomy. This of course opened up a lot of potentially interesting challenges to further investigate in future related works.

\section*{Acknowledgements}
This research has received funding from the European Union’s Horizon 2020 research and innovation program under grant No. 957218 (\textit{IntellIoT}) and from the Swiss National Science Foundation under grant No. 189474 (\textit{HyperAgents}).

\bibliographystyle{splncs04}
\bibliography{bibliography}

\end{document}